\documentclass{aa}
\usepackage{graphics}
\usepackage{times}

\begin{document}

\title{Astrometric radial velocities}
\subtitle{I. Non-spectroscopic methods for measuring stellar radial velocity
\thanks{Based (in part) on observations by the ESA Hipparcos satellite}}

\thesaurus{11(03.13.2; 03.20.7; 05.01.1; 08.04.1; 08.11.1; 10.15.1)}

\author{ 
   Dainis Dravins,
   Lennart Lindegren, and
   S{\o}ren Madsen
} 

\institute{
   Lund Observatory, Box~43, 
   SE--22100 Lund, Sweden
(e-mail: 
dainis@astro.lu.se, 
lennart@astro.lu.se, 
soren@astro.lu.se)
}

\date{Received 11 February 1999; accepted 26 May 1999}

\titlerunning{Astrometric radial velocities I. Non-spectroscopic methods}
\authorrunning{Dravins et al.}
\maketitle

\begin{abstract}

High-accuracy astrometry permits the determination of not only stellar tangential 
motion, but also the component along the line-of-sight.  Such non-spectroscopic 
(i.e.\ astrometric) radial velocities are independent of stellar atmospheric 
dynamics, spectral complexity and variability, as well as of gravitational 
redshift.  Three methods are analysed: (1) changing annual parallax, (2) changing 
proper motion and (3) changing angular extent of a moving group of stars.  
All three have significant potential in planned astrometric projects.  Current 
accuracies are still inadequate for the first method, while the second is 
marginally feasible and is here applied to 16 stars. The third method reaches 
high accuracy ($<1$~km~s$^{-1}$) already with present data, although for some
clusters an accuracy limit is set by uncertainties in the cluster expansion
rate.


\end{abstract}

\section{Introduction}

For well over a century, radial velocities for objects outside the solar system 
have been determined through spectroscopy, using the (Doppler) shifts of stellar 
spectral lines.  The advent of high-accuracy (sub-milliarcsec) astrometric 
measurements, both on ground and in space, now permits radial velocities to be 
obtained by alternative methods, based on geometric principles and therefore 
independent of spectroscopy. The importance of such {\it astrometric radial 
velocities\/} stems from the fact that they are independent of phenomena 
which affect the spectroscopic method, such as line asymmetries and shifts 
caused by atmospheric pulsation, surface convection, stellar rotation, stellar 
winds, isotopic composition, pressure, and gravitational potential.  Conversely, the 
differences between spectroscopic and astrometric radial velocities may provide 
information on these phenomena that cannot be obtained by other methods.  
Although the theoretical possibility of deducing astrometric radial velocities 
from geometric projection effects was noted already at the beginning of the 
20th century (if not earlier), it is only recently that such methods have 
reached an accuracy level permitting non-trivial comparison with spectroscopic 
measurements.

We have analysed three methods by which astrometric radial velocities can be 
determined (Fig.~\ref{fig:methods}).  Two of them are applicable to individual, 
nearby stars and are based on the well understood secular changes in the stellar 
trigonometric parallax and proper motion.  The third method uses the apparent 
changes in the geometry of a star cluster or association to derive its kinematic 
parameters, assuming that the member stars share, in the mean, a common space 
velocity.  In Sects.~\ref{sec:pidot} to \ref{sec:mcm} we describe the principle 
and underlying assumptions of each of the three methods and derive approximate 
formulae for the expected accuracy of resulting astrometric radial velocities.  
For the first and second methods, an inventory of nearby potential target stars 
is made, and the second method is applied to several of these.

However, given currently available astrometric data, only the third 
(moving-cluster) method is capable of yielding astrophysically interesting, 
sub-km~s$^{-1}$ accuracy. In subsequent papers we develop in detail the theory 
of this method, based on the maximum-likelihood principle, as well as its 
practical implementation, and apply it to a number of nearby open clusters 
and associations, using data from the Hipparcos astrometry satellite.

\section{Notations}
\label{sec:not}

In the following sections, $\pi$, $\mu$ and $v_r$ denote the trigonometric 
parallax of a star, its (total) proper motion, and its radial velocity.  The 
components of $\mu$ in right ascension and declination are denoted 
$\mu_{\alpha *}$ and $\mu_{\delta}$, with  
$\mu=(\mu_{\alpha *}^2+\mu_{\delta}^2)^{1/2}$.  The dot signifies a time 
derivative, as in $\dot\pi \equiv {\rm d}\pi/{\rm d}t$.  
The statistical uncertainty 
(standard error) of a quantity $x$ is denoted $\epsilon(x)$.  (We prefer this 
non-standard notation to $\epsilon_x$, since $x$ is itself often a subscripted 
variable.)  $\sigma_v$ is used for the physical velocity dispersion in a 
cluster.  $A=1.49598\times 10^8$~km is the astronomical unit; the equivalent 
values $4.74047$~km~yr~s$^{-1}$ and $9.77792\times 10^8$~mas~km~yr~s$^{-1}$
are conveniently used in equations below (cf.\ Table~1.2.2 in Vol.~1 of ESA \cite{esa}). 
Other notations are explained as they are introduced.

\section{Astrometric accuracies}
\label{sec:acc}

In estimating the potential accuracy of the different methods, we consider three 
hypothetical situations: 
\begin{itemize}
\item{}
Case A: a quasi-continuous series of observations over a few years, resulting in 
an accuracy of $\epsilon(\pi)=1$~mas (milliarcsec) for the trigonometric  
parallaxes and $\epsilon(\mu)=1$~mas~yr$^{-1}$ for the proper motions.  
\item{}
Case B: similar to Case~A, only a thousand times better, i.e.\ 
$\epsilon(\pi)=1$~$\mu$as (microarcsec) and $\epsilon(\mu)=1$~$\mu$as~yr$^{-1}$.
\item{}
Case C: {\it two\/} sets of measurements, separated by an interval of 50~yr, 
where each set has the same accuracy as in Case~B.  The much longer-time 
baseline obviously allows a much improved determination of the accumulated 
changes in parallax and proper motion.  
\end{itemize}

The accuracies assumed in Case~A are close to what the Hipparcos space  
astrometry mission (ESA \cite{esa}) achieved for its main observation programme of 
more than 100$\,$000 stars.  Current ground-based proper motions may be slightly 
better than this, but not by a large factor.  This case therefore represents, 
more or less, the state-of-the-art accuracy in optical astrometry.  Accuracies 
in the 1 to 10~$\mu$as range are envisaged for some planned or projected 
space astrometry missions, such as GAIA 
(Lindegren \& Perryman \cite{lindegren96}) and SIM (Unwin et al.\ \cite{unwin}).
The duration of such a mission is here assumed to be about 5~years.   
Using the longer-time baselines 
available with ground-based techniques, similar performance may in the future be 
reached with the most accurate ground-based techniques (Pravdo \& Shaklan 
\cite{pravdo}; Shao \cite{shao}).  
Case~B therefore corresponds to what we could realistically hope 
for within one or two decades.  Case~C, finally, probably represents an upper 
limit to what is practically feasible in terms of long-term proper-motion 
accuracy, not to mention the patience of astronomers.

\begin{figure}[!t]
  \resizebox{\hsize}{!}{\includegraphics*{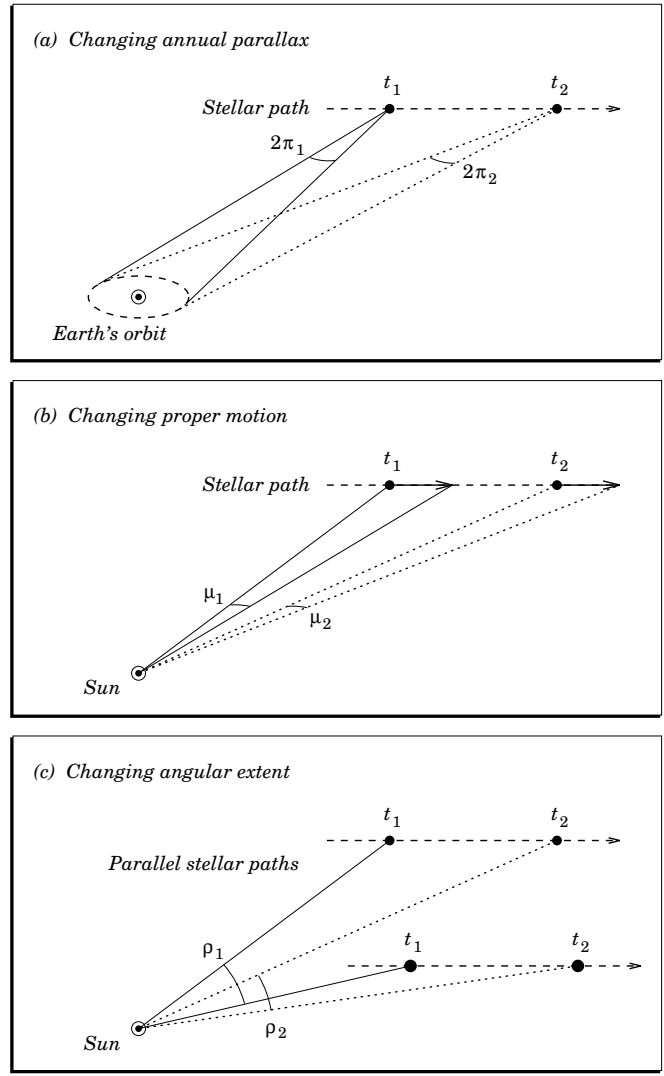}}
  \caption[ ]{Three methods to astrometrically determine stellar radial motion: 
{\bf a} Changing trigonometric parallax $\pi$; {\bf b} Perspective change in the 
proper motion $\mu$; {\bf c} Changing angular separation $\rho$ of stars 
sharing the same space velocity, e.g.\ in a moving  cluster.}\label{fig:methods}
\end{figure}

\section{Radial velocity from changing annual parallax}
\label{sec:pidot}

The most direct and model-independent way to determine radial velocity by 
astrometry is to measure the secular change in the trigonometric parallax 
(Fig.~\ref{fig:methods}a).  The distance $b$ (from the solar system barycentre) 
is related to parallax $\pi$ through $b=A/\sin\pi\simeq A/\pi$.  Since 
$v_r=\dot{b}$, the radial velocity is 
\begin{equation}
  v_r=-A\,\frac{\dot\pi}{\pi^2}
  \, ,  \label{eq:pidot}
\end{equation}
where $A$ is the astronomical unit (Sect.~\ref{sec:not}).  The 
equivalent of Eq.~(\ref{eq:pidot}) was derived by Schlesinger (\cite{schlesinger}), 
who concluded that the parallax change is very small for every known star.  However, 
although extremely accurate parallax measurements are obviously required, the 
method is not as unrealistic as it may seem at first.  To take a specific, if 
extreme, example: for Barnard's star (Gl~699~=~HIP~87937), with $\pi=549$~mas 
and $v_r=-110$~km~s$^{-1}$, the expected parallax rate is 
$\dot\pi=+34~\mu$as~yr$^{-1}$.  According to our discussion in  
Sect.~\ref{sec:acc} this will almost certainly be measurable in the near future.

It can be noted that the changing-parallax method, in contrast to the methods 
described in Sect.~\ref{sec:mudot} and \ref{sec:mcm}, does not depend on the 
object having a large and uniform space motion, and would therefore be 
applicable to all stars within a few parsecs of the Sun.

\begin{table}[t]
\caption[ ]{Number of target stars, and achievable accuracy in the radial 
velocity, as obtained from the changing annual parallax. $n$~=~number of stars 
with parallax greater than $\pi$ in the Third Catalogue of Nearby Stars (Gliese 
\& Jahrei{\ss} \cite{gliese}); $\epsilon(v_r)$~=~predicted accuracy according to  
Eqs.~(\ref{eq:epspidotA}) and (\ref{eq:epspidotB}). Case~B: space astrometry 
mission lasting $L=5$~yr and yielding parallaxes with standard error 
$\epsilon(\pi)=1~\mu$as and consequently 
$\epsilon(\dot\pi)=0.69~\mu$as~yr$^{-1}$; Case~C: 
combination of two such measurements, 
$\epsilon(\pi_1)=\epsilon(\pi_2)=1~\mu$as, with a large epoch difference,  
$T=50$~yr.}
\begin{flushleft}
\begin{tabular}{lrcc}
\hline\noalign{\smallskip}
$\pi$ & $n$ &
\multicolumn{2}{c}{$\epsilon(v_r)$~[km~s$^{-1}$]} \\
 {}[mas] && Case~B & Case~C \\
\noalign{\smallskip}
\hline
\noalign{\smallskip}
740 &   3 & \phantom{0}1.2\phantom{} &  0.05\phantom{} \\
300 &  14 & \phantom{0}7.5\phantom{} &  0.3\phantom{0} \\
200 &  60 & \phantom{}17\phantom{.0} &  0.7\phantom{0} \\
100 & 326 & \phantom{}68\phantom{.0} &  2.8\phantom{0} \\
\noalign{\smallskip}
\hline
\end{tabular}
\end{flushleft}\label{tab:epspidot}
\end{table}

\subsection{Achievable accuracy}
\label{sec:epspidot}

The accuracy in $v_r$ is readily estimated from Eq.~(\ref{eq:pidot}) for a given 
accuracy in $\dot\pi$, since the contribution of the parallax uncertainty to the 
factor $A/\pi^2$ is negligible.  The achievable accuracy in $\dot\pi$ depends 
both on the individual astrometric measurements and on their number and  
distribution in time.  Concerning the temporal distribution of the measurements 
we consider two limiting situations:

{\it Quasi-continuous observation.\/}  The measurements are more or less 
uniformly spread out over a time period of length $L$ centred on the epoch 
$t_0$.  This is a good approximation to the way a single space mission would 
typically be operated; for example, Hipparcos had $L \simeq 3$~yr and $t_0 
\simeq 1991.25$.  In such a case there exist simple (mean) relations between how 
accurately the different astrometric parameters of the same star can be derived, 
depending on $L$.  For instance, $\epsilon(\mu_\delta)\simeq  
(\sqrt{12}/L)\epsilon(\delta)$, if $\epsilon(\mu_\delta)$ is the accuracy of the 
proper motion in declination and $\epsilon(\delta)$ that of the declination at 
$t_0$.  This approximation is applicable to Case~A and B as defined in 
Sect.~\ref{sec:acc}.

{\it Two-epoch observation.\/}  Two isolated parallax or proper-motion 
measurements are taken, separated by a long time interval (say, $T$ years) 
during which no observation takes place.  Each measurement must actually be the 
result of a series covering at least a year or so, but the duration of each such 
series is assumed to be negligible compared with $T$.  This could be two similar 
space missions separated by several decades and is applicable to Case~C in 
Sect.~\ref{sec:acc}.

For quasi-continuous observation we may assume that the parallax variation is 
linear over the observation period $L$.  Thus, $\pi(t)=\pi_0+(t-t_0)\dot\pi$, 
where $\pi_0$ and $\dot\pi$ are two parameters to be determined from the 
observations. There exists an approximate relation between the accuracies of 
these two parameters that is similar to that between the proper motion and the 
position at the mean epoch, viz.\ $\epsilon(\dot\pi) =  
(\sqrt{12}/L)\epsilon(\pi_0)$. Moreover, the estimates of the two parameters are 
uncorrelated, so $\epsilon(\pi_0)$ equals the accuracy $\epsilon(\pi)$ of a 
parallax determination in the absence of the parallax-change term; thus
\begin{equation}
  \epsilon(v_r) \simeq
  \sqrt{12}\,A\,\frac{\epsilon(\pi)}{L\pi^2} \, .
  \label{eq:epspidotA}
\end{equation}
In the case of a two-epoch observation, let us assume that independent parallax 
measurements $\pi_1$ and $\pi_2$ are made at epochs $t_1$ and $t_2=t_1+T$.  The 
estimated rate of change is $\dot\pi=(\pi_2-\pi_1)/T$.  With $\epsilon(\pi_1)$ 
and $\epsilon(\pi_2)$ denoting the accuracies of the two measurements, we have
$\epsilon(\dot\pi)=[\epsilon(\pi_1)^2+\epsilon(\pi_2)^2]^{1/2}/T$ and  
consequently
\begin{equation}
  \epsilon(v_r) \simeq \frac{A}{T\pi^2}\,
  \left[\epsilon(\pi_1)^2+\epsilon(\pi_2)^2\right]^{1/2} \, .
  \label{eq:epspidotB}
\end{equation}
For given observational errors we find, from both Eq.~(\ref{eq:epspidotA}) and 
(\ref{eq:epspidotB}), that the radial-velocity error is simply a function of 
distance. The number of potential target stars for a certain maximum 
radial-velocity uncertainty is therefore given by the total number of stars 
within the 
corresponding maximum distance. Table~\ref{tab:epspidot} gives the actual 
numbers of such stars, and the observational accuracies that may be reached.

\section{Radial velocity from changing proper motion
(perspective acceleration)}\label{sec:mudot}

To a good approximation, single stars move with uniform linear velocity through 
space.  For a given linear tangential velocity, the angular velocity (or proper 
motion $\mu$), as seen from the Sun, varies inversely with the distance to the 
object.  However, the tangential velocity changes due to the varying angle  
between the line of sight and the space-velocity vector  
(Fig.~\ref{fig:methods}b).  As is well known (e.g.\ van de Kamp \cite{vdkamp67}, 
Murray \cite{murray}) the two effects combine to produce an apparent 
(perspective) acceleration 
of the motion on the sky, or a rate of change in proper motion amounting to 
$\dot\mu = -2\mu v_r/b$.  With $b=A/\pi$ we find 
\begin{equation}
  v_r = -A\,\frac{\dot\mu}{2\pi\mu} \, .
  \label{eq:mudot}
\end{equation}
Schlesinger (\cite{schlesinger}) 
derived the equivalent of this equation, calculated the 
perspective acceleration for Kapteyn's and Barnard's stars (cf.\ 
Table~\ref{tab:epsmudot}) and noted that, if accurate positions are acquired 
over long periods of time, ``we shall be in position to determine the radial 
velocities of these stars independently of the spectroscope and with an 
excellent degree of precision''.  The equation for the perspective acceleration 
was earlier derived by Seeliger (\cite{seeliger})\footnote{Some remarks in the
literature, e.g.\ by Ristenpart (\cite{ristenpart}) and Lundmark \& Luyten 
(\cite{lundmark}), seem to suggest that the perspective acceleration was
discovered by Bessel (\cite{bessel44}).  However, as far as we can determine, 
Bessel only discussed proper-motion changes caused by gravitational 
perturbations, explicitly neglecting terms depending on the radial motion.}
and used by Ristenpart (\cite{ristenpart}) in an 
(unsuccessful) attempt to determine $\dot\mu$ observationally for  
Groombridge~1830.  A major consideration for Ristenpart seems to have been the 
possibility to derive the parallax from the apparent acceleration in combination 
with a spectroscopic radial velocity. Such a determination of `acceleration 
parallaxes' was also considered by Eichhorn (\cite{eichhorn81}).

\begin{table*}[t!]
\caption[ ]{Nearby high-proper-motion stars suitable for the determination of 
radial velocity from changing proper motion (perspective acceleration). All 
known objects with a parallax-proper-motion product greater than  
0.5~arcsec$^2$~yr$^{-1}$ are included.  Data are from the Hipparcos Catalogue 
(ESA \cite{esa}) where available, otherwise from the preliminary version of the Third 
Catalogue of Nearby Stars, CNS3 (Gliese \& Jahrei{\ss} \cite{gliese}). Columns:  
CNS3~=~identifier in CNS3; HD~=~identifier in the HD/HDE catalogue; 
HIP~=~identifier in the Hipparcos Catalogue; Sp~=~spectral classification in 
CNS3; $V$~=~visual magnitude; $b$~=~distance from the Sun; $\pi\mu$~=~product of 
parallax and proper motion; $\epsilon(v_r)$~=~predicted accuracy of the 
astrometric radial velocity (Case~B: space astrometry mission lasting $L=5$~yr 
and yielding an accuracy of $\epsilon(\mu)=1~\mu$as~yr$^{-1}$ for the mean 
proper motion and consequently $\epsilon(\dot\mu)=1.55~\mu$as~yr$^{-2}$ for the 
acceleration; Case~C: the combination of two such measurements with 
$\epsilon(\mu_1)=\epsilon(\mu_2)=1~\mu$as~yr$^{-1}$ having an epoch difference 
$T=50$~yr). In the last column, $P$ is the orbital period of a binary. Possibly, 
the star CF~UMa does not exist, being an erroneous  catalogue entry.
}
\begin{flushleft}
\begin{tabular}[h]{lrrlrccccl}
\hline\noalign{\smallskip}
CNS3 & HD~~ & HIP~~ & Sp & $V$~ & $b$ & $\pi\mu$ & 
\multicolumn{2}{c}{$\epsilon(v_r)$~[km~s$^{-1}$]} & Remark \\
&&&&& [pc] & [arcsec$^2$~yr$^{-1}$] & Case~B & Case~C \\ 
\noalign{\smallskip}
\hline
\noalign{\smallskip}
Gl~699   &        &  87937 & sdM4   &  9.5 & 1.8 & 5.69 &0.13&0.01&
Barnard's star \\ 
Gl~551   &        &  70890 & M5Ve   & 11.0 & 1.3 & 2.98 &0.25&0.01&
$\alpha$~Cen~C (Proxima) \\
Gl~559B  & 128621 &  71681 & K0V    &  1.4 & 1.3 & 2.76 &0.27&0.01&
$\alpha$~Cen~B \\
Gl~559A  & 128620 &  71683 & G2V    &  0.0 & 1.3 & 2.75 &0.28&0.01&
$\alpha$~Cen~A (AB: $P=80$~yr) \\
Gl~191   &  33793 &  24186 & M0V    &  8.9 & 3.9 & 2.21 &0.34&0.01&
Kapteyn's star \\
Gl~887   & 217987 & 114046 & M2Ve   &  7.4 & 3.3 & 2.10 &0.36&0.01&
\\
Gl~406   &        &        & M6     & 13.5 & 2.4 & 1.96 &0.38&0.01&
Wolf 359 \\
Gl~411   &  95735 &  54035 & M2Ve   &  7.5 & 2.5 & 1.88 &0.40&0.01&
\\
Gl~820A  & 201091 & 104214 & K5Ve   &  5.2 & 3.5 & 1.52 &0.50&0.02&
61~Cyg~A \\
Gl~820B  & 201092 & 104217 & K7Ve   &  6.1 & 3.5 & 1.48 &0.51&0.02&
61~Cyg~B (AB: $P=700$~yr) \\
Gl~1     & 225213 &    439 & M4V    &  8.6 & 4.4 & 1.40 &0.54&0.02&
\\
Gl~845   & 209100 & 108870 & K5Ve   &  4.7 & 3.6 & 1.30 &0.58&0.02&
$\epsilon$~Ind \\
Gl~65A   &        &        & dM5.5e & 12.6 & 2.6 & 1.28 &0.59&0.02&
\\
Gl~65B   &        &        & dM5.5e & 12.7 & 2.6 & 1.28 &0.59&0.02&
UV~Cet (AB: $P=27$~yr) \\
Gl~273   &        &  36208 & M3.5   &  9.8 & 3.8 & 0.98 &0.77&0.03&
Luyten's star \\
Gl~866AB &        &        & M5e    & 12.3 & 3.4 & 0.96 &0.79&0.03&
$P=2.2$~yr \\
Gl~412A  &        &  54211 & M2Ve   &  8.8 & 4.8 & 0.93 &0.81&0.03&
\\
Gl~412B  &        &        & M6e    & 14.4 & 5.3 & 0.86 &0.88&0.03&
WX~UMa \\
Gl~825   & 202560 & 105090 & M0Ve   &  6.7 & 3.9 & 0.88 &0.86&0.03&
\\
Gl~15A   &   1326 &   1475 & M2V    &  8.1 & 3.6 & 0.82 &0.92&0.03&
GX~And \\
Gl~15B   &        &        & M6Ve   & 11.1 & 3.6 & 0.84 &0.90&0.03&
GQ~And (AB: $P=2600$~yr)\\
Gl~166A  &  26965 &  19849 & K1Ve   &  4.4 & 5.0 & 0.81 &0.94&0.03&
40~Eri~A ($o^2$~Eri) \\
Gl~166B  &  26976 &        & DA4    &  9.5 & 4.8 & 0.84 &0.90&0.03&
40~Eri~B (BC: $P=250$~yr)\\
Gl~166C  &        &        & dM4.5e &  9.5 & 4.8 & 0.84 &0.90&0.03&
DY~Eri \\
Gl~299   &        &        & dM5    & 12.8 & 6.8 & 0.77 &0.98&0.04&
Ross 619 \\
Gl~451A  & 103095 &  57939 & G8VI   &  6.4 & 9.2 & 0.77 &0.98&0.04&
Groombridge~1830 \\
Gl~451B  &        &   & ---& 12\phantom{.0}& 9.2 & 0.82 &0.92&0.03&
CF~UMa (non-existent star?)\\
Gl~35    &        &   3829 & DZ7    & 12.4 & 4.4 & 0.68 &1.1\phantom{0}&0.04&
van Maanen~2 \\
Gl~725B  & 173740 &  91772 & dM5    &  9.7 & 3.6 & 0.66 &1.2\phantom{0}&0.04&
AB: $P=400$~yr\\
Gl~725A  & 173739 &  91768 & dM4    &  8.9 & 3.6 & 0.63 &1.2\phantom{0}&0.04&
\\
Gl~440   &        &  57367 & DQ6    & 11.5 & 4.6 & 0.58 &1.3\phantom{0}&0.05&
\\
Gl~71    &  10700 &   8102 & G8Vp   &  3.5 & 3.6 & 0.53 &1.4\phantom{0}&0.05&
$\tau$~Cet\\
Gl~754   &        &        & M4.5   & 12.2 & 5.7 & 0.52 &1.5\phantom{0}&0.05&
\\
Gl~139   &  20794 &  15510 & G5V    &  4.3 & 6.1 & 0.52 &1.5\phantom{0}&0.05&
82~Eri\\
Gl~905   &        &        & dM6    & 12.3 & 3.2 & 0.51 &1.5\phantom{0}&0.05&
\\
Gl~244A  &  48915 &  32349 & A1V    &$-1.4$& 2.6 & 0.51 &1.5\phantom{0}&0.05&
$\alpha$~CMa (Sirius)\\
Gl~244B  &        &        & DA2    &      & 2.6 & 0.51 &1.5\phantom{0}&0.05&
AB: $P=50$~yr\\
Gl~53A   &   6582 &   5336 & G5VI   &  5.2 & 7.6 & 0.50 &1.5\phantom{0}&0.06&
$\mu$~Cas\\
Gl~53B   &        &   & ---& 11\phantom{.0}& 7.6 & 0.51 &1.5\phantom{0}&0.06&
AB: $P=20$~yr\\
\noalign{\smallskip}
\hline
\end{tabular}
\end{flushleft}\label{tab:epsmudot}
\end{table*}

Subsequent attempts to determine the perspective acceleration of Barnard's star
by Lundmark \& Luyten (\cite{lundmark}), Alden (\cite{alden}) and van de Kamp 
(\cite{vdkamp35b}) yielded 
results that were only barely significant or (in retrospect) spurious.
Meanwhile, Russell \& Atkinson (\cite{russell}) suggested that the white dwarf 
van Maanen~2 might exhibit a gravitational redshift of several hundred
km~s$^{-1}$ and that this could be distinguished from a real radial velocity
through measurement of the perspective acceleration.  The astrophysical
relevance of astrometric radial-velocity determinations was thus already
established (Oort \cite{oort}).

In relatively recent times, the perspective acceleration was successfully 
determined for Barnard's star by van de Kamp (\cite{vdkamp62}, \cite{vdkamp63}, 
\cite{vdkamp67}, \cite{vdkamp70}, \cite{vdkamp81}); for van Maanen~2 by 
van de Kamp (\cite{vdkamp71}), Gatewood \& Russell (\cite{gatewood}) and 
Hershey (\cite{hershey}); and for Groombridge 1830 by 
Beardsley et al.\ (\cite{beardsley}).  
Among these determinations the highest precisions, in terms of the astrometric 
radial velocity, were obtained for Barnard's star (corresponding to 
$\pm 4$~km~s$^{-1}$; van de Kamp \cite{vdkamp81}) and van Maanen~2 
($\pm 15$~km~s$^{-1}$; Gatewood \& Russell \cite{gatewood}).

Our application of the method, combining Hipparcos measurements with data in the 
Astrographic Catalogue, yielded radial velocities for 16 objects, as listed in 
Table~\ref{tab:AC}.

\subsection{Achievable accuracy}
\label{sec:epsmudot}

The accuracy of the radial velocity calculated from Eq.~(\ref{eq:mudot}) can be 
estimated as in Sect.~\ref{sec:epspidot}. It depends on the 
parallax-proper-motion product $\pi\mu$.  
The most promising targets for this method are listed 
in Table~\ref{tab:epsmudot}, which contains the known nearby stars ranked after 
decreasing $\pi\mu$. 

For quasi-continuous observation during a period of length $L$ we may use a 
quadratic model for the angular position $\phi$ of the star along the 
great-circle arc: 
$\phi(t) = \phi(t_0) + (t-t_0)\mu_0 + {\scriptstyle\frac{1}{2}} (t-t_0)^2\dot\mu$.  
Here $\mu_0$ is the proper motion at the central epoch $t_0$. 
The estimates of $\mu_0$ and $\dot\mu$ are found to be uncorrelated and their 
errors related by $\epsilon(\dot\mu) = (\sqrt{60}/L)\epsilon(\mu_0)$.  
Consequently,
\begin{equation}
  \epsilon(v_r) \simeq \sqrt{15}\,A\,\frac{\epsilon(\mu)}{L\pi\mu}
  \label{eq:epsmudotA}
\end{equation}
where $\epsilon(\mu)$ is the accuracy of proper-motion measurements in the 
absence of temporal changes.  We neglect the (small) contribution to 
$\epsilon(v_r)$ from the uncertainty in the denominator $\pi\mu$.

For a two-epoch observation, consider proper-motion measurements $\mu_1$ and 
$\mu_2$ made around $t_1$ and $t_2=t_1+T$.  The estimated acceleration is 
$\dot\mu=(\mu_2-\mu_1)/T$. Provided the two observation intervals centred on 
$t_1$ and $t_2$ do not overlap, the measurements are independent, yielding the 
standard error  
$\epsilon(\dot\mu)=[\epsilon(\mu_1)^2+\epsilon(\mu_2)^2]^{1/2}/T$.
For the radial velocity this gives
\begin{equation}
  \epsilon(v_r) \simeq \frac{A}{T\pi\mu}\,
  \left[\epsilon(\mu_1)^2+\epsilon(\mu_2)^2\right]^{1/2} \, .
  \label{eq:epsmudotB}
\end{equation}
Based on these formulae, Table~\ref{tab:epsmudot} gives the potential 
radial-velocity accuracy for the two cases B and C defined in Sect.~\ref{sec:acc}.

In a two-epoch observation we normally have, in addition, a very good estimate 
of the {\it mean\/} proper motion between $t_1$ and $t_2$, provided the 
positions $\phi_1$ and $\phi_2$ at these epochs are accurately known.  In the 
previous quadratic model we may take the reference epoch to be $t_0=(t_1+t_2)/2$ 
and find $\mu_0=(\phi_2-\phi_1)/T$ with standard error  
$\epsilon(\mu_0)=[\epsilon(\phi_1)^2+\epsilon(\phi_2)^2]^{1/2}/T$. The three 
proper-motion estimates $\mu_0$, $\mu_1$ and $\mu_2$ (referred to $t_0$, $t_1$ 
and $t_2$) are mutually independent and may be combined in a least-squares 
estimate of $\dot\mu$. If $\epsilon(\mu_1)=\epsilon(\mu_2)$ (equal weight at 
$t_1$ and $t_2$), then it is found that $\mu_0$ does not contribute at all to 
the determination of $\dot\mu$, and the standard error is still given by 
Eq.~(\ref{eq:epsmudotB}).  If, on the other hand, the two observation epochs are 
not equivalent, then some improvement can be expected by introducing the 
position measurements.

An important special case is when there is just a position (no proper motion) 
determined at one of the epochs, say $t_1$. This is however equivalent to the 
two independent proper-motion determinations $\mu_0$ at $t_0=(t_1+t_2)/2$, and 
$\mu_2$ at $t_2$, separated by $t_2-t_0=T/2$. Applying Eq.~(\ref{eq:epsmudotB}) 
on this case yields
\begin{equation}
  \epsilon(v_r) \simeq \frac{2A}{T\pi\mu}\,
  \left[\frac{\epsilon(\phi_1)^2+\epsilon(\phi_2)^2}{T^2}+
        \epsilon(\mu_2)^2\right]^{1/2} \, .
  \label{eq:epsmudotC}
\end{equation}
This formula is applicable on the combination of a recent position and 
proper-motion measurement (e.g.\ by Hipparcos) with a position derived from old  
photographic plates (e.g.\ the Astrographic Catalogue). Taking $t_1\sim 1907$, 
$\epsilon(\phi_1)\simeq 200$~mas as representative for the Astrographic  
Catalogue, and $t_2=1991.25$, $\epsilon(\phi_2)\simeq 1$~mas,  
$\epsilon(\mu_2)\simeq 1$~mas~yr$^{-1}$ for Hipparcos, we find  
$\epsilon(v_r)\simeq (60~\mbox{km~s$^{-1}$~arcsec$^2$~yr$^{-1}$})/(\pi\mu)$.  
With such data, moderate accuracies of a few tens of km~s$^{-1}$ can be reached 
for several stars (Sect.~\ref{sec:pm-results}).

\subsection{Effects of gravitational perturbations}

The perspective-acceleration method depends critically on the assumption that
the star moves with uniform space motion relative the observer.  The presence
of a real acceleration of their relative motions, caused by gravitational 
action of other bodies, would bias the calculated astrometric radial velocity 
by $-g_t/2\mu$, where $g_t$ is the tangential component of the relative 
acceleration.  The acceleration towards the Galactic centre caused by the 
smoothed Galactic potential in the vicinity of the Sun is 
$g \simeq 2\times 10^{-13}$~km~s$^{-2}$.  For a hypothetical observer near
the Sun but unaffected by this acceleration, the maximum bias would 
be 0.06~km~s$^{-1}$ for Barnard's star, and 0.17~km~s$^{-2}$ for Proxima.
However, since real observations are made relative the solar-system
barycentre, which itself is accelerated in the Galactic gravitational field, 
the observed (differential) effect will be very much smaller.  

In the case of Proxima the 
acceleration towards $\alpha$~Cen~AB is of a similar magnitude as the Galactic 
acceleration.  For the several orbital binaries in Table~\ref{tab:epsmudot} the 
curvature of the orbit is much greater than the perspective acceleration. 
Application of this method will therefore require careful correction for all 
known perturbations: the possible presence of long-period companions may 
introduce a considerable uncertainty.

Among other effects which may have to be considered are light-time effects 
which, to first order in $c^{-1}$, may require a correction of $-v_t^2/2c$ on 
the right-hand side of Eq.~(\ref{eq:mudot}), where $v_t$ is the tangential 
velocity.  For typical high-velocity (Population~II) stars the correction is 
0.1--0.2~km~s$^{-1}$.  At this accuracy level, the precise definition 
of the radial-velocity concept itself requires careful consideration (Lindegren 
et al.\ \cite{lindegren99}).

\begin{table*}[t]
\caption[ ]{Astrometric radial velocities, obtained by combining positions and 
proper motions from Hipparcos (epoch 1991.25) with old position measurements 
from the Astrographic Catalogue. Spectroscopic radial velocities are also given 
(from Turon et al.\ \cite{turon}). Data for the binary 
61~Cyg = HIP~104214+104217 = AC~1382645+1382649 
refer to its mass centre, assuming a mass ratio of 0.90. Two 
solutions are given: the first based on Hipparcos and AC data alone; the second 
(marked *) includes Bessel's visual measurements from 1838. }
\label{tab:AC}
\begin{flushleft}
\begin{tabular}{llcccl}
\hline\noalign{\smallskip}
~~HIP & \multicolumn{2}{c}{AC~2000} & 
  \multicolumn{2}{c}{Radial velocity, $v_r$ [km~s$^{-1}$]} & Remark \\
    & ~~~No. & Epoch   &  Astrom. & Spectr. & \\
\noalign{\smallskip}
\hline
\noalign{\smallskip}
\phantom{000}439 & 3152964 & 1912.956 & \phantom{00}$+7.0~\pm\phantom{0}29.7$ 
  & \phantom{0}$+22.9$ & \\
\phantom{00}1475 & 1406215 & 1898.435 &  \phantom{0}$-40.8~\pm\phantom{0}36.1$ 
  & \phantom{0}$+12.0$ & GX~And \\
\phantom{00}5336 & 1721511 & 1913.868 & $-106.4~\pm\phantom{0}82.7$ 
  & \phantom{0}$-98.0$ & $\mu$~Cas \\
\phantom{0}15510 & 3488626 & 1901.018 &  \phantom{0}$+89.6~\pm\phantom{0}59.1$ 
  & \phantom{0}$+86.8$ & 82~Eri \\
\phantom{0}19849 & 2125614 & 1892.970 &  \phantom{0}$-73.2~\pm\phantom{0}32.8$ 
  & \phantom{0}$-42.6$ & 40~Eri \\
\phantom{0}24186 & 3505363 & 1899.058 & $+249.4~\pm\phantom{0}13.7$ 
  &$+245.5$ & Kapteyn's star \\
\phantom{0}36208 & \phantom{0}282902 & 1908.859 &\phantom{0}$-
33.7~\pm\phantom{0}38.4$ 
  & \phantom{0}$+18.7$ & Luyten's star \\
\phantom{0}54035 & 1340883 & 1930.895 &  \phantom{0}$+22.0~\pm\phantom{0}36.1$ 
  & \phantom{0}$-85.6$ & \\
\phantom{0}54211 & 1463341 & 1895.620 &  \phantom{0}$+58.6~\pm\phantom{0}30.5$ 
  & \phantom{0}$+67.5$ & \\
\phantom{0}57367 & 4195112 & 1924.492 &  \phantom{0}$+43.1~\pm\phantom{}106.4$ 
  &  --- & \\
\phantom{0}57939 & 1342199 & 1930.260 & $-139.5~\pm\phantom{0}86.1$ 
  & \phantom{0}$-98.0$ & Groombridge 1830 \\
\phantom{0}87937 & \phantom{0}146626 & 1905.979 & $-101.9~\pm\phantom{00}6.5$ 
  &$-109.7$ & Barnard's star \\
\phantom{}104214/217 & 1382645/649 & 1921.699 & \phantom{0}$-
12.2~\pm\phantom{0}34.5$  
  & \phantom{0}$-64.5$ & 61~Cyg \\
\phantom{}104214/217 & 1382645/649 & 1921.699 
  & \phantom{0}$-68.0~\pm\phantom{0}11.1$\rlap{~*} & ~~'' & ~~~~'' \\
\phantom{}105090 & 3462277 & 1905.316 & \phantom{0}$-25.0~\pm\phantom{0}40.9$ 
  & \phantom{0}$+23.6$ \\
\phantom{}108870 & 4384302 & 1901.189 & \phantom{0}$-64.1~\pm\phantom{0}23.6$ 
  & \phantom{0}$-40.0$ & $\epsilon$~Ind \\
\phantom{}114046 & 3355101 & 1913.368 & \phantom{0}$+54.0~\pm\phantom{0}19.7$ 
  & \phantom{00}$+9.5$ & \\
\noalign{\smallskip}
\hline
\end{tabular}
\end{flushleft}
\end{table*}

\subsection{Results from observed proper-motion changes}
\label{sec:pm-results}

Past determinations of the perspective acceleration, e.g.\ by van de Kamp 
(\cite{vdkamp81}) 
and Gatewood \& Russell (\cite{gatewood}), were based on photographic observations 
collected over several decades, in which the motion of the target star was 
measured relative to several background (reference) stars.  One difficulty with 
the method has been that the positions and motions of the reference stars are 
themselves not accurately known, and that small errors in the reference data  
could cause a spurious acceleration of the target star 
(van de Kamp \cite{vdkamp35a}).

The Hipparcos Catalogue (ESA \cite{esa}) established a very accurate and homogeneous 
positional reference frame over the whole sky.  Using the proper motions, this 
reference frame can be extrapolated backwards in time.  It is then possible to 
re-reduce measurements of old photographic plates, and express even century-old 
stellar positions in the same reference frame as modern observations.  This 
should greatly facilitate the determination of effects such as the perspective  
acceleration, which are sensitive to systematic errors in the reference frame.

As part of the {\it Carte du Ciel\/} project begun more than a century ago, an 
astrographic programme to measure the positions of all stars down to the 11th 
magnitude was carried out and published as the Astrographic Catalogue, AC (see 
Eichhorn \cite{eichhorn74} for a description).  After transfer to electronic media, the 
position measurements have been reduced to the Hipparcos reference frame  
(Nesterov et al.\ \cite{nesterov}, Urban et al.\ \cite{urban}).  
The result is a positional 
catalogue of more than 4~million stars with a mean epoch around 1907 and a 
typical accuracy of about 200~mas.  We have used the version known as AC~2000 
(Urban et al.\ \cite{urban}), available on CD-ROM from the US Naval Observatory, 
to examine the old positions of all the stars with HIP identifiers in  
Table~\ref{tab:epsmudot}.  

For the stars in Table~\ref{tab:AC} we successfully matched the AC positions 
with the positions extrapolated backwards from the Hipparcos Catalogue and hence 
could calculate the astrometric radial velocities. Other potential targets in 
Table~\ref{tab:epsmudot} were either outside the magnitude range of AC~2000 
(e.g.\ $\alpha$~Cen and Proxima) or lacked an accurate 
proper motion from Hipparcos (e.g.\ van Maanen~2 and HIP~91768+91772).

The basic procedure was as follows.  The rigorous epoch transformation algorithm 
described in Sect.~1.5.5 of the Hipparcos Catalogue, Vol.~1, was used to  
propagate the Hipparcos position and its covariance matrix to the AC~2000 epoch 
relevant for each star.  This extrapolated position was compared with the actual 
measured position in AC~2000, assuming a standard error of 200~mas in each 
coordinate for the latter.  A $\chi^2$ goodness-of-fit was then calculated from 
the position difference and the combined covariance of the extrapolated and 
measured positions. The epoch transformation algorithm requires that the radial 
velocity is known. The radial velocity was therefore varied until the $\chi^2$ 
attained its minimum value. The $\pm$1$\sigma$ confidence interval given in the 
table was obtained by modifying the radial velocity until the $\chi^2$ had 
increased by one unit above the minimum.

\begin{figure}[tb]
  \resizebox{\hsize}{!}{\includegraphics*{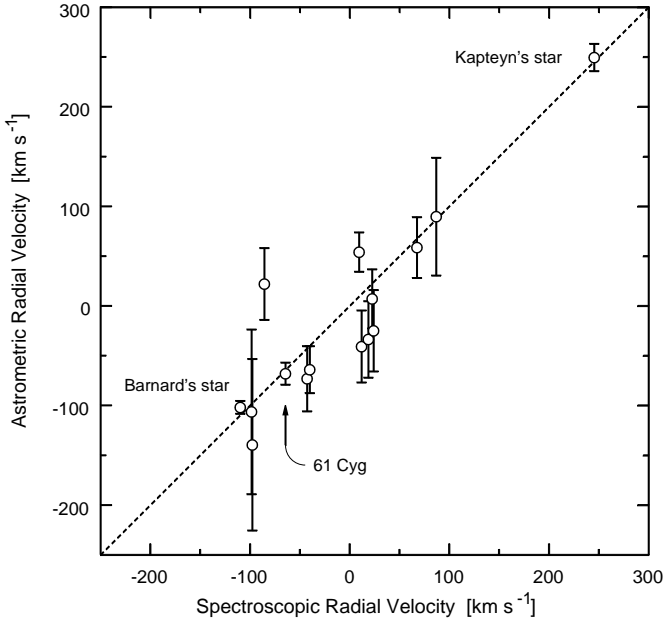}}
  \caption[ ]{Comparison of the astrometric radial velocities in Table~\ref{tab:AC} 
  with their spectroscopic counterparts. The straight line is the expected 
  relationship $v_r(\mbox{astrom.})\simeq v_r(\mbox{spectr.})$.
  The three most accurate determinations are indicated.}
\label{fig:res}
\end{figure}

For some of the stars, data had to be corrected for duplicity or known orbital 
motion.  The solutions for the resolved binary 61~Cyg (HIP~104214+104217) refer 
to the mass centre, assuming a mass ratio of $M_{\rm B}/M_{\rm A}=0.90$, as 
estimated by means of standard isochrones from the absolute magnitudes and 
colour indices of the components (S{\"o}derhjelm, private communication).  For 
the astrometric binary $\mu$~Cas (HIP~5336) the Hipparcos data explicitly refer 
to the mass centre using the orbit by Heintz \& Cantor (\cite{heintz}); the same 
orbit was used to correct the AC position of the primary to the mass centre. 
No correction for orbital motion was used for GX~And and 40~Eri.

Table~\ref{tab:AC} gives two solutions for 61~Cyg.  The first solution was  
obtained as described above, using only the Hipparcos data plus the AC positions 
for the two components.  The second solution, marked with an asterisk in the 
table, was derived by including also the observations by Bessel (\cite{bessel39}) 
from his 
pioneering determination of the star's parallax. Bessel measured the angular 
distances from the geometrical centre (half-way between the components) 
of 61~Cyg to two reference stars, called 
$a$ and $b$ in his paper.  After elimination of aberration, proper motion and 
parallax, he found the distances $461.6171 \pm 0.015$~arcsec and $706.2791 \pm 
0.017$~arcsec for the beginning of year 1838 (B1838.0 = J1838.0022).  The 
uncertainties are our estimates (standard errors) based on the scatter of the 
residuals in Bessel's solution `II'.  We identified the reference stars in 
AC~2000 and in the Tycho Catalogue (ESA \cite{esa}) as $a$ = AC~1382543 = 
TYC~3168~708~1 and $b$ = AC~1382712 = TYC~3168~1106~1.  Extrapolating the 
positions from these catalogues back to B1838 allowed us to compute the position 
of the geometrical centre of 61~Cyg in the Hipparcos/Tycho reference frame.  This 
could then be transformed to the position of the mass centre, using Bessel's own 
measurement of the separation and position angle in 61~Cyg and the previously 
assumed mass ratio.  Actually, all the available data were combined into a 
$\chi^2$ goodness-of-fit measure and the radial velocity was varied in order to 
find the minimum and the $\pm$1$\sigma$ confidence interval.  This gave $v_r=-
68.0 \pm 11.1$~km~s$^{-1}$.

Table~\ref{tab:AC} also gives the spectroscopic radial velocities when available 
in the literature.  A comparison between the astrometric and spectroscopic 
radial velocities is made in Fig.~\ref{fig:res}.  Given the stated confidence 
intervals, the agreement is in all cases rather satisfactory.  The exercise 
demonstrates the basic feasibility of this method, but also hints at some of the 
difficulties in applying it to non-single stars.

\section{Radial velocity from changing angular extent
(moving-cluster method)}\label{sec:mcm}

The moving-cluster method is based on the assumption that the stars in a cluster 
move through space essentially with a common velocity vector. The 
radial-velocity component makes the cluster appear to contract or expand due to 
its changing distance (Fig.~\ref{fig:methods}c).  The relative rate of apparent 
contraction equals the relative rate of change in distance to the cluster.  This 
can be converted to a linear velocity (in km~s$^{-1}$) if the distance to the 
cluster is known, e.g.\ from trigonometric parallaxes.  In practice, the method  
amounts to determining the space velocity of the cluster, i.e.\ the convergent 
point and the speed of motion, through a combination of proper motion and 
parallax data.  Once the space velocity is known, the radial velocity for any
member star may be calculated by projecting the velocity vector onto the
line of sight.

The method can be regarded as an inversion of the classical procedure 
(e.g.\ Binney \& Merrifield \cite{binney}) by which the distances to the stars
in a moving cluster are derived from the proper motions and
(spectroscopic) radial velocities: if instead the distances are known, the
radial velocities follow.  The first application of the classical 
moving-cluster method for distance determination was by Klinkerfues 
(\cite{klinkerfues}), in a study of the Ursa Major system.  The possibility to 
check spectroscopic radial velocities against astrometric data was recognised
by Klinkerfues, but could not then be applied to the Ursa Major cluster due to 
the lack of reliable trigonometric parallaxes.  This changed with
Hertzsprung's (\cite{hertzsprung}) discovery that Sirius probably belongs to 
the Ursa Major moving group.  The relatively large and well-determined parallax 
of Sirius, combined with its considerable angular distance from the cluster apex, 
could lead to a meaningful estimate for the cluster velocity and hence for 
the radial velocities.  Rasmuson (\cite{rasmuson}) and Smart (\cite{smart}) 
appear to have been among the first who actually made this computation,
although mainly as a means of verifying the cluster method for distance
determination.  Later studies by Petrie (\cite{petrie49}) and Petrie \& Moyls 
(\cite{petrie53}) reached formal errors in the astrometric radial velocities 
below 1~km~s$^{-1}$.  The last paper concluded ``There does not appear to be 
much likelihood of improving the present results until a substantial 
improvement in the accuracy of the trigonometric parallaxes becomes possible.''

One of the purposes of the Petrie \& Moyls study was to derive the astrometric 
radial velocities of spectral type A in order to check the Victoria system of 
spectroscopic velocities. The method was also applied to the Hyades (Petrie 
\cite{petrie63}) but only with an uncertainty of a few km~s$^{-1}$.  Given the 
expected future availability of more accurate proper motions and trigonometric  
parallaxes, Petrie (\cite{petrie62}) envisaged that one or two moving clusters 
could eventually be used as primary radial-velocity standards for early-type 
spectra.

Such astrometric data are now in fact available.  In Sect.~\ref{sec:3prec} we 
derive a rough estimate of the accuracy of the method and survey nearby clusters 
and associations in order to find promising targets for its application.  
An important consideration is to what extent systematic velocity patterns in 
the cluster, in particular cluster expansion, will limit the achievable accuracy.  
This is discussed in Sect.~\ref{sec:3syst} and Appendix~A.
In Sect.~\ref{sec:indpar} we briefly consider the improvement in the distance 
estimates for individual stars resulting from the moving-cluster method.

The present discussion of the moving-cluster method is only intended to 
highlight its theoretical potential and limitations. Its actual application 
requires a more rigorous formulation, which is developed in a second paper.

\begin{table*}[t]
\caption[ ]{Potential accuracy in astrometric radial velocities of nearby star 
clusters and associations, using the moving-cluster method. The first five
data columns contain general cluster data taken from the literature (see
below): $n=$~number of member stars; (photometric) ages; 
$\rho_{\rm rms}=$~rms angular radius; $b_0=$~mean distance from the Sun; 
$v_r$, $v_t=$~approximate radial and tangential velocities.  The columns headed 
$\epsilon(v_{0r})$ give the potential accuracies according to Eq.~(\ref{eq:epsmc1}) 
assuming $\epsilon(\pi)=1$~mas and $\epsilon(\mu)=1$~mas~yr$^{-1}$ (Case~A), or 
$\epsilon(\pi)=1~\mu$as and $\epsilon(\mu)=1~\mu$as~yr$^{-1}$ (Case~B).
The final column gives the bias in the astrometric radial velocity estimate
(Eq.~\ref{eq:deltaexp}) that would result from neglecting an isotropic expansion 
of the cluster, assuming that the relative expansion rate equals the inverse age 
of the cluster.  A dash means that no data were available.
The cluster data are mostly from Lyng{\aa} (\cite{lynga87a}, \cite{lynga87b}) 
[assuming $\rho_{\rm rms}$ to equal the cluster core diameter $D$ in that catalogue], 
with supplementary data from de~Zeeuw et al.\ (\cite{dezeeuw}), 
de~Zeeuw \& Brand (\cite{dezeeuw+brand}),
Mermilliod et al.\ (\cite{mermilliod}), Perryman et al.\ (\cite{perryman}),
Smith \& Messer (\cite{smith}), Soderblom \& Mayor (\cite{soderblom}), 
and Stryker \& Hrivnak (\cite{stryker}).
Tangential velocities
and the angular radii of the larger clusters and of all the associations  
were calculated from data for identified members in the Hipparcos 
Catalogue (ESA \cite{esa}).  The memberships for the associations were 
adopted from de~Zeeuw et al.\ (\cite{dezeeuw}) and 
Platais et al.\ (\cite{platais}).  Ursa~Major refers to the 
rather extended group (UMaG) identified by Soderblom \& Mayor (\cite{soderblom});  
`HIP~98321' to the possible association identified by 
Platais et al.\ (\cite{platais}).
An internal velocity dispersion $\sigma_v=0.25$~km~s$^{-1}$ was assumed for all 
clusters and associations, although a higher value is more likely for associations 
and the halos of some clusters.}
\begin{flushleft}
\begin{tabular}{lcccccccccccc}
\hline\noalign{\smallskip}
Name & IAU & $n$ & Age & $\rho_{\rm rms}$ & $b_0$ & $v_r$ & $v_t$ &~~~&
\multicolumn{2}{c}{$\epsilon(v_{0r})$~[km~s$^{-1}$]} && 
$\delta_{\rm exp}(v_{0r})$ \\
& designation && [Myr] & [arcmin] & [pc] &\multicolumn{2}{c}{[km~s$^{-1}$]}
&& (A) & (B) && [km~s$^{-1}$] \\
\noalign{\smallskip}
\hline
\noalign{\smallskip}
Cassiopeia--Taurus & & \phantom{00}83 & \phantom{00}25 & \phantom{}1800 & 
\phantom{0}190 & \phantom{0}$+$6&21&&
\phantom{0}0.24\phantom{} & 0.06 &&\phantom{00}$-$7.3\phantom{0}\\
Upper Centaurus Lupus & & \phantom{0}221 & \phantom{00}13 & \phantom{0}670 & 
\phantom{0}140 & \phantom{0}$+$5&21&&
\phantom{0}0.25\phantom{} & 0.09\phantom{} &&\phantom{0}$-$10\phantom{.00} \\
Ursa Major & & \phantom{00}40 & \phantom{0}300 & \phantom{}4300 & 
\phantom{00}25 & $-$11&\phantom{0}5&&
\phantom{0}0.11\phantom{} & 0.10\phantom{} &&\phantom{00}$-$0.08\\
Lower Centaurus Crux & & \phantom{0}180 & \phantom{00}10 & \phantom{0}560 & 
\phantom{0}118 & $+$12&19&&
\phantom{0}0.30\phantom{} & 0.12\phantom{} &&\phantom{0}$-$11\phantom{.00}\\
Hyades & C~0424$+$157 & \phantom{0}380 & \phantom{0}625 & \phantom{0}560 & 
\phantom{00}46 & $+$43&25&&
\phantom{0}0.19\phantom{} & 0.14\phantom{} &&\phantom{00}$-$0.07\\
Perseus OB3 ($\alpha$~Per) & C~0318$+$484 & \phantom{0}186 & \phantom{00}50 &
\phantom{0}350 & \phantom{0}180 & \phantom{0}$-$1&29
&&\phantom{0}0.64\phantom{} & 0.18\phantom{} &&\phantom{00}$-$3.4\phantom{0}\\
`HIP~98321' & & \phantom{00}59 & \phantom{00}60 & \phantom{0}740 & 
\phantom{0}300 & $-$17&\phantom{0}4
&&\phantom{0}1.1\phantom{0} & 0.19 &&\phantom{00}$-$4.8\phantom{0}\\
Upper Scorpius & & \phantom{0}120 & \phantom{000}5 & \phantom{0}325 & 
\phantom{0}145 & \phantom{0}$-$5&18&&
\phantom{0}0.71\phantom{} & 0.24\phantom{} &&\phantom{0}$-$28\phantom{.00}\\
Lacerta OB1 & & \phantom{00}96 & \phantom{00}16 & \phantom{0}350 & 
\phantom{0}370 & $-$13&\phantom{0}8
&&\phantom{0}1.8\phantom{0} & 0.26\phantom{} &&\phantom{0}$-$22\phantom{.00}\\
Collinder 121 & & \phantom{0}103 & \phantom{000}5 & \phantom{0}290 & 
\phantom{0}540 & $+$26&15&&
\phantom{0}3.3\phantom{0} & 0.32\phantom{} &&\phantom{}$-$103\phantom{.00}\\
Collinder 70 & C~0533$-$011 & \phantom{0}345 & \phantom{}--- & \phantom{0}140 & 
\phantom{0}430 & ---&---&&
\phantom{0}2.7\phantom{0} & 0.33\phantom{} &&---\\
Cepheus OB2 & & \phantom{00}71 & \phantom{000}5 & \phantom{0}320 & 
\phantom{0}615 & $-$21&12&&
\phantom{0}4.1\phantom{0} & 0.34\phantom{} &&\phantom{}$-$120\phantom{.00}\\
Vela OB2 & & \phantom{00}93 & \phantom{00}20 & \phantom{0}260 & 
\phantom{0}415 & $+$18&20&&
\phantom{0}2.8\phantom{0} & 0.36\phantom{} &&\phantom{0}$-$20\phantom{.00}\\
Perseus OB2 & & \phantom{00}41 & \phantom{000}7 & \phantom{0}340 & 
\phantom{0}300 & $+$20&14&&
\phantom{0}2.5\phantom{0} & 0.43\phantom{} &&\phantom{}$-$46\phantom{.00}\\
Pleiades & C~0344$+$239 & \phantom{0}277 & \phantom{0}130 & \phantom{0}120 & 
\phantom{0}125 & \phantom{0}$+$7&29&&
\phantom{0}1.1\phantom{0} & 0.43\phantom{} &&\phantom{00}$-$0.92\\
Coma Berenices & C~1222$+$263 & \phantom{0}273 & \phantom{0}460 & \phantom{0}120 & 
\phantom{00}87 & \phantom{0$+$}0&\phantom{0}9&&
\phantom{0}0.84\phantom{} & 0.43\phantom{} &&\phantom{00}$-$0.18\\
NGC 3532 & C~1104$-$584 & \phantom{0}677 & \phantom{0}290 & \phantom{00}50 & 
\phantom{0}480 & \phantom{0}$+$7&27&&
\phantom{0}6.0\phantom{0} & 0.66\phantom{} &&\phantom{00}$-$1.6\phantom{0}\\
Praesepe & C~0837$+$201 & \phantom{0}161 & \phantom{0}830 & \phantom{00}70 & 
\phantom{0}160 & $+$33&26&&
\phantom{0}3.1\phantom{0} & 0.98\phantom{} &&\phantom{00}$-$0.18\\
NGC 2477 & C~0750$-$384 & \phantom{}1911 & \phantom{}1260 & \phantom{00}20 & 
\phantom{}1150 &\phantom{0}$+$7&17&&
\phantom{}21\phantom{.00} & 0.98\phantom{} &&\phantom{00}$-$0.87\\
IC 4756 & C~1836$+$054 & \phantom{0}466 & \phantom{0}830 & \phantom{00}39 & 
\phantom{0}390 & $-$18&\phantom{0}5&&
\phantom{0}7.6\phantom{0} & 1.0\phantom{0} &&\phantom{00}$-$0.45\\
IC 4725 & C~1828$-$192 & \phantom{0}601 & \phantom{00}41 & \phantom{00}29 & 
\phantom{0}710 &\phantom{0}$+$3&20&&
\phantom{}16\phantom{.00} & 1.2\phantom{0} &&\phantom{0}$-$17\phantom{.00}\\
Trumpler 10 & & \phantom{00}23 & \phantom{00}15 & \phantom{0}150 & 
\phantom{0}370 & $+$21&26&&
\phantom{0}8.6\phantom{0} & 1.2\phantom{0} &&\phantom{0}$-$24\phantom{.00}\\
Cepheus OB6 & & \phantom{00}20 & \phantom{00}50 & \phantom{0}150 & 
\phantom{0}270 & $-$20&22&&
\phantom{0}6.8\phantom{} & 1.3\phantom{0} &&\phantom{00}$-$5.2\phantom{0}\\
NGC 752 & C~0154$+$374 & \phantom{00}77 & \phantom{}3300 & \phantom{00}75 & 
\phantom{0}360 & \phantom{0}$-$3&20&&
\phantom{0}9.0\phantom{0} & 1.3\phantom{0} &&\phantom{00}$-$0.10\\
NGC 6618 & C~1817$-$162 & \phantom{0}660 & \phantom{}--- & \phantom{00}25 & 
\phantom{}1500 & ---&26&&
\phantom{}38\phantom{.00} & 1.3\phantom{0} &&---\\
NGC 2451 & C~0743$-$378 & \phantom{0}153 & \phantom{00}41 & \phantom{00}50 & 
\phantom{0}315 & $+$27&27&&
\phantom{0}8.4\phantom{0} & 1.4\phantom{0} &&\phantom{00}$-$7.3\phantom{0}\\
NGC 7789 & C~2354$+$564 & \phantom{0}583 & \phantom{}1780 & \phantom{00}25 & 
\phantom{}1800 &$-$54&27&&
\phantom{}49\phantom{.00} & 1.4\phantom{0} &&\phantom{00}$-$1.0\phantom{0}\\
NGC 2099 & C~0549$+$325 & \phantom{}1842 & \phantom{0}200 & \phantom{00}14 & 
\phantom{}1300 & \phantom{0}$+$8&45&&
\phantom{}35\phantom{.00} & 1.4\phantom{0} &&\phantom{00}$-$6.4\phantom{0}\\
NGC 6475 & C~1750$-$348 & \phantom{00}54 & \phantom{0}130 & \phantom{00}80 & 
\phantom{0}240 & $-$12&\phantom{0}7&&
\phantom{0}6.8\phantom{0} & 1.5\phantom{0} &&\phantom{00}$-$1.8\phantom{0}\\
NGC 2264 & C~0638$+$099 & \phantom{0}222 & \phantom{00}10 & \phantom{00}39 & 
\phantom{0}800 & $+$22&17&&
\phantom{}23\phantom{.00} & 1.5\phantom{0} &&\phantom{0}$-$76\phantom{.00}\\
Stock 2 & C~0211$+$590 & \phantom{0}166 & \phantom{0}170 & \phantom{00}45 & 
\phantom{0}300 & \phantom{0}$+$2&30&&
\phantom{0}8.6\phantom{0} & 1.5\phantom{0} &&\phantom{00}$-$1.7\phantom{0}\\
IC 2602 & C~1041$-$641 & \phantom{00}33 & \phantom{00}29 & \phantom{0}100 & 
\phantom{0}150 & $+$22&14&&
\phantom{0}4.6\phantom{0} & 1.5\phantom{0} &&\phantom{00}$-$5.0\phantom{0}\\
\noalign{\smallskip}
\hline
\end{tabular}
\end{flushleft}\label{tab:clus}
\end{table*}

\subsection{Potential accuracy}\label{sec:3prec}

The accuracy of the astrometric radial velocity potentially achievable by 
the moving-cluster 
method can be estimated as follows.  Let $b$ be the (mean) distance to the 
cluster and consider a star at angular distance $\rho$ from the centre of  the 
cluster, as seen from the Sun.  The projected linear distance of the star from 
the centre of the cluster is $b\sin\rho \simeq b\rho$, provided the angular 
extent of the cluster is not very large.  As the cluster moves through space, 
its linear dimensions remain constant, so that $\dot\rho b + \rho\dot b=0$.  
Putting $\dot\rho=\mu$ (the proper motion relative to the cluster centre), 
$\dot b = v_r$, and $b=A/\pi$, gives $v_r=-A\mu/(\rho\pi)$.  Now suppose that 
the parallaxes and proper motions of $n$ cluster stars are measured, each with 
uncertainties of $\epsilon(\pi)$ and $\epsilon(\mu)$.  Standard error 
propagation formulae give the expected accuracy in $v_r$ as
\begin{equation}
  \epsilon(v_r) \simeq
  A\,\frac{\epsilon(\mu)}{\rho_{\rm rms}\pi\sqrt{n}}\,
  \left[ 1 + \left(\frac{v_r\rho_{\rm rms}\epsilon(\pi)}
  {A\epsilon(\mu)}\right)^2 \right]^{1/2} \, 
  \label{eq:epsmc0}
\end{equation}
where $\rho_{\rm rms}$ is in radians; $A$ is the astronomical unit 
(Sect.~\ref{sec:not}).  The expression within the square brackets derives 
from the uncertainty in the mean cluster distance, by which the derived radial 
velocity scales.  For the type of (space) astrometry data considered here 
(Case~A and B), $\epsilon(\pi)/\epsilon(\mu)$ is on the order of a few years  
(for Hipparcos the mean ratio is $\simeq 1.2$~yr). The factor in brackets can 
then be neglected except for the most extended (and nearby) clusters.

Under certain circumstances it is not the accuracy of proper-motion measurements 
that defines the ultimate limit on $\epsilon(v_r)$, but rather internal velocity 
dispersion among the cluster stars. Assuming isotropic dispersion with standard 
deviation $\sigma_v$ in each coordinate, one must add $\sigma_v\pi/A$ 
quadratically to the measurement error $\epsilon(\mu)$ in  
Eq.~(\ref{eq:epsmc0}).  Thus 
\begin{eqnarray}
  \epsilon(v_{0r}) &\simeq& \frac{1}{\rho_{\rm rms}\sqrt{n}}\,
  \left[\sigma_v^2+\left(\frac{A\epsilon(\mu)}{\pi}\right)^2
  \right]^{1/2} \nonumber \\
  && \times\left[ 1 + \left(\frac{v_r\rho_{\rm rms}\epsilon(\pi)}
  {A\epsilon(\mu)}\right)^2 \right]^{1/2} \, ,
  \label{eq:epsmc1}
\end{eqnarray}
is the accuracy achievable for the radial velocity of the cluster centroid.  For 
the radial velocity of an individual star this uncertainty must be increased by 
the internal dispersion.

The internal velocity dispersion will dominate the error budget for nearby 
clusters, viz.\ if $\pi>A\epsilon(\mu)/\sigma_v$. Assuming a velocity 
dispersion of 0.25~km~s$^{-1}$ and a proper-motion accuracy of 1~mas~yr$^{-1}$ 
(as for Hipparcos), this will be the case for clusters within 50~pc of the Sun.  
For an observational accuracy in the 1--10~$\mu$as~yr$^{-1}$ range the internal 
dispersion will dominate in practically all Galactic clusters and  
Eq.~(\ref{eq:epsmc1}) can be simplified to $\epsilon(v_{0r}) \simeq 
\sigma_v/(\rho_{\rm rms}\sqrt{n})$. In this case the achievable accuracy 
becomes independent of the astrometric one.

Table~\ref{tab:clus} lists some nearby clusters and associations, with estimates 
of the achievable accuracy in the radial velocity of the cluster centroid, 
assuming current (Hipparcos-type) astrometric performance (Case~A in 
Sect.~\ref{sec:acc}) as well as future (microarcsec) expectations (Case~B).  
As explained above, increasing the astrometric accuracy still further gives 
practically no improvement; this is why Case~C is not considered in the table.

The entry `HIP~98321' refers to the possible association identified by 
Platais et al.\ (\cite{platais}) and named after one of its members.  
Of dubious status, it was included as an example of the extended, 
low-density groups that may exist in the general stellar field, but 
are difficult to identify with existing data.

\subsection{Internal velocity fields, including cluster expansion}
\label{sec:3syst}

Blaauw (\cite{blaauw64}) showed that the proper motion pattern for a linearly 
expanding cluster is identical to the apparent convergence produced by 
parallel space motions.  Astrometric data alone therefore cannot distinguish
such expansion from a radial motion.  If such an expansion exists, and is not 
taken into account in estimating the astrometric radial velocity, a bias will
result, as examined in Appendix~A (Eq.~\ref{eq:bias}).  

The gravitationally unbound associations are
known to expand on timescales comparable with their nuclear ages 
(de~Zeeuw \& Brand \cite{dezeeuw+brand}).  But also for a gravitationally 
bound open cluster some expansion can be expected as a result of the 
dynamical evolution of the cluster (see Mathieu \cite{mathieu85} and 
Wielen \cite{wielen88} for an introduction to this complex issue).  
In either case the inverse age of the cluster or association may be 
taken as a rough upper limit on the cluster's relative expansion 
rate $\kappa$ [yr$^{-1}$].
Eq.~(\ref{eq:bias}) then gives
\begin{equation}
  |\delta_{\rm exp}(v_{0r})| \la
  0.9543 \times \frac{b_0~[{\rm pc}]}{\rm age~[Myr]}~{\rm km}~{\rm s}^{-1}\, 
  \label{eq:deltaexp}
\end{equation}
for the bias of a star near the centre of the cluster.  (For an expanding
cluster, $\delta_{\rm exp}$ is always negative.)  
Resulting values, in the last column of Table~\ref{tab:clus}, 
are adequately small for a few nearby, relatively old clusters.  In other cases
the potential bias is very large and will certainly limit the applicability 
of the method.  The OB associations are particularly troublesome, not only 
because they are young objects (implying large values of $\kappa$), but also 
because they sometimes appear to expand significantly faster than their 
photometric ages would suggest (de~Zeeuw \& Brand \cite{dezeeuw+brand}).

However, it should be remembered that the ultimate limitation set by cluster
expansion depends on how accurately the expansion rate $\kappa$ can be estimated 
by some independent means.  For instance, if $\kappa$ can somehow be estimated 
to within 10~per cent of its value, then the residual biases would still be 
on the sub-km~s$^{-1}$ level for most of the objects in Table~\ref{tab:clus}.
Numerical simulation of the dynamical evolution of clusters might in
principle provide such estimates of $\kappa$, as could the spectroscopic radial 
velocities as function of distance.  The use of spectroscopic data would 
not necessarily defeat the purpose of the method, i.e.\ to determine 
{\it absolute\/} radial velocities, since the expansion is revealed already 
by {\it relative\/} measurements.

\subsection{Distances to the individual stars}
\label{sec:indpar}

In a rigorous estimation of the space motion of a moving cluster, such
as will be presented in a second paper, the distances to the individual 
member stars of the cluster appear as parameters to be estimated.  
A by-product of the method is therefore that the individual distances are 
improved, sometimes considerably, compared with the original trigonometric 
distances (Dravins et al.\ \cite{dravins}; Madsen \cite{madsen}).  The 
improvement results from a combination of the trigonometric parallax 
$\pi_{\rm trig}$ with the kinematic (secular) parallax 
$\pi_{\rm kin}=A\mu/v_t$ derived from the star's proper motion $\mu$ and 
tangential velocity $v_t$, the latter obtained from the estimated 
space velocity vector of the cluster.  The accuracy of the combined
parallax estimate $\widehat{\pi}$ can be estimated from 
$\epsilon(\widehat{\pi})^{-2}=\epsilon(\pi_{\rm trig})^{-2}+
\epsilon(\pi_{\rm kin})^{-2}$.  In calculating $\epsilon(\pi_{\rm kin})$ we 
need to take into account the observational uncertainty in $\mu$ and the
uncertainty in $v_t$ from the internal velocity dispersion.  The result is
\begin{equation}
  \epsilon(\widehat{\pi}) \simeq 
  \left[ \epsilon(\pi_{\rm trig})^{-2} +                   \frac{(v_t/A)^2}{\epsilon(\mu)^2+(\pi\sigma_v/A)^2}
  \right]^{-1/2} \, .
  \label{eq:epspihat}
\end{equation}
From the data in Table~\ref{tab:clus} we find that, in Case~A, the 
moving-cluster method will be useful to resolve the depth structures of the 
Hyades cluster and of the associations Cassiopeia--Taurus, Upper Centaurus 
Lupus, Lower Centaurus Crux, Perseus OB3 and Upper Scorpius.  In Case~B, 
all the clusters and associations are resolved by the trigonometric 
parallaxes, so the kinematic parallaxes will bring virtually no improvement.

Calculation of kinematic distances to stars in moving clusters is 
of course a classical procedure (e.g., Klinkerfues \cite{klinkerfues} 
and van Bueren \cite{vanbueren}); what is new in our treatment is that such 
distances are derived without recourse to spectroscopic data.

\section{Further astrometric methods}
\label{sec:moremethods}

With improved astrometric data, further methods for radial-velocity 
determinations may become feasible.  The moving-cluster method could in 
principle be applied to any geometrical configuration of a fixed linear 
size.
To reach an accuracy of 1~km~s$^{-1}$ in the astrometric radial velocity
of an object at 10~pc distance requires a dimensional `stability' of 
the order of $10^{-7}$~yr$^{-1}$; at a distance of 1~kpc the requirement
is $10^{-9}$~yr$^{-1}$.  Since these numbers are greater than or 
comparable with the inverse dynamical timescales of many types of 
galactic objects, there is at least a theoretical chance that the
method could work, given a sufficient measurement accuracy.  We consider
briefly two such possibilities.

\subsection{Binary stars}

According to the previous argument it would be possible to ignore the
relative motion of the components in a binary if the period is
longer than some $10^7$~yr.  This implies an linear separation of at 
least some 50\,000~astronomical units, or a few degrees on the sky at 
10~pc distance.  In principle, then, this case is equivalent to a moving 
cluster with $n=2$ stars.

In the opposite case of a (relatively) short-period binary, the radial 
velocity might be obtained from apparent changes of the orbit.  The 
projected orbit will not be closed, but form a spiral on the sky: slightly 
diverging if the stars are approaching, slightly converging if they recede.  
For a system at a distance of 10~pc, say, with a component separation of 
10~astronomical units, a radial velocity of 100~km~s$^{-1}$ will change 
the apparent orbital radius of 1~arcsec by 10~$\mu$as per year.
The relative positions would need to be measured during at least a significant 
fraction of an orbital period, or some 20~years in our example, resulting 
in an accumulated change by about 0.2~mas.

Since only relative position measurements between the same stars are required, 
the observational challenges are not as severe as in some other cases.  For a  
binary with a few arcsec separation, the isoplanatic properties of the 
atmosphere imply that the cross-correlation distance between the speckle images 
of both stars should be stable to better than one mas. Averaging very many 
exposures should reduce the errors into the $\mu$as range, with practical 
limits possibly set by differential refraction (McAlister \cite{mcalister}).

\subsection{Globular clusters}

The moving-cluster method (Sect.~\ref{sec:mcm}) could in principle be applied
also to globular clusters.  Globular clusters differ markedly from open 
clusters in that (potentially) many more stars could be measured, and through
a much larger velocity dispersion ($\sim 5$~km~s$^{-1}$; Peterson \& King 
\cite{peterson}).  The higher
number of stars partly compensates the larger dispersion.  However, all
globular clusters are rather distant, making their angular radii small.
As discussed in Sect.~\ref{sec:3prec} the approximate formula
$\epsilon(v_{0r}) \simeq \sigma_v/(\rho_{\rm rms}\sqrt{n})$ applies in
the case when the internal motions are well resolved.  Taking
$\rho_{\rm rms} \sim 10$--20~arcmin as representative for the angular
radii of comparatively nearby globular clusters, we find that averaging
over some $3\times 10^4$ to $10^5$ member stars is needed to reach a 
radial-velocity accuracy comparable with $\sigma_v$, i.e.\ several 
km~s$^{-1}$.
Furthermore, in view of the discussion in Sect.~\ref{sec:3syst}, it is
not unlikely that the complex kinematic structures of these objects
(e.g.\ Norris et al.\ \cite{norris}) would bias the results.  Thus, globular 
clusters remain difficult targets for astrometric radial-velocity 
determination.

\section{Conclusions}

The theoretical possibility to use astrometric data (parallax and proper motion) 
to deduce the radial motions of stars has long been recognised. With the highly 
accurate (sub-milliarcsec) astrometry already available or foreseen in planned 
space missions, such radial-velocity determinations are now also a practical 
possibility.  This will have implications for the future definition of 
radial-velocity standards, as the range of geometrically determined 
accurate radial velocities, 
hitherto limited to the solar system and to solar-type spectra, is extended to 
many other stellar types represented in the solar neighbourhood.

We have identified and analysed three main methods to determine astrometric 
radial velocities. The first method, using the changing annual parallax, is the 
intuitively most obvious one, but requires data of an accuracy beyond current 
techniques.  It is nevertheless potentially interesting in view of future space 
missions or long-term observations from the ground.

The second method, using the changing proper motion or perspective acceleration 
of stars, has a long history, and was previously applied to a few objects, 
albeit with modest precision in the resulting radial velocity.  New results for 
a greater number of stars, obtained by combining old and modern data, were given 
in Table~\ref{tab:AC} and Fig.~\ref{fig:res}, thus proving the concept.  
However, to realise the full potential of the method again requires the 
accuracies of future astrometry projects.

In both these methods the uncertainty in the astrometric radial velocity  
increases, statistically, with distance squared.  They are therefore in practice 
limited to relatively few stars very close to the Sun and, in the second method, 
to stars with a large tangential velocity. In the general case, the two methods 
could actually be combined to yield a somewhat higher accuracy, but at least for 
the stars considered in Tables~\ref{tab:epspidot} and \ref{tab:epsmudot} this 
would only bring a marginal improvement.

The third method, using the changing angular extent of a moving cluster or 
association, is an inversion of the classical moving-cluster method, 
by which the distance to the cluster was derived from its radial velocity and 
convergence point.  If the distance is known from trigonometric parallaxes, one 
can instead calculate the radial velocities.  It appears to be the only method 
by which astrophysically interesting accuracies can be obtained with 
existing astrometric data.  In future papers we will develop and exploit this 
possibility in full, using data from the Hipparcos mission.  A by-product 
of the method is that the distance estimates to individual cluster 
stars may be significantly improved compared with the parallax 
measurements.

One would perhaps expect the moving-cluster method to become extremely powerful 
with the much more accurate data expected from future astrometry projects.  
Unfortunately, this is not really the case, as internal velocities (both random 
and systematic) become a limiting factor as soon as they are resolved by the 
proper-motion measurements.  Nevertheless, even the limited number of clusters 
within reach of such determinations contain a great many stars spanning a wide 
range in spectral type and luminosity.

\begin{acknowledgements}

This project is supported by the Swedish National Space Board and the Swedish 
Natural Science Research Council.  We thank Dr.~S.~S{\"o}derhjelm (Lund 
Observatory) for providing information on double and multiple stars,
Prof.~P.T.~de Zeeuw and co-workers (Leiden Observatory) for advance data on 
nearby OB associations, and the referee, Prof.\ A.~Blaauw, for stimulating
criticism of the manuscript.

\end{acknowledgements}

\appendix

\section{Effects of internal velocities on the moving-cluster velocity estimate}

In this Appendix we examine how sensitive the moving-cluster method is
to {\it systematic\/} velocity patterns in the cluster, and to what extent
such patterns can be determined independently of the astrometric radial
velocity.  For this purpose we may ignore the random motions as well as 
the observational errors and we consider only a linear (first-order)
velocity field.

Let $\vec{b}_0$ be the position of the cluster centroid relative to the Sun and 
$\vec{s}=\vec{b}-\vec{b}_0$ the position of a member star relative to the 
centroid.  The space velocity of the star is $\vec{v}=\vec{v}_0+\vec{\eta}$, 
where $\vec{\eta}$ is the peculiar velocity.  The velocity field is described 
by the tensor $\vec{T}$ such that 
$\vec{\eta}=\vec{T}\vec{s}$. In Cartesian coordinates the components of this 
tensor are simply the partial derivatives $\partial v_\alpha/\partial b_\beta$ 
for $\alpha,\beta=x,y,z$.  
These nine numbers together describe the three components of a 
rigid-body rotation, three components of an anisotropic expansion or 
contraction, and three components of linear shear.

It is intuitively clear that certain components of the linear velocity field, 
such as rotation about the line of sight, can be determined purely from the 
astrometric data.  If the corresponding components of $\vec{T}$ are included as 
parameters in the cluster model, they can be estimated and will not produce a 
systematic error (bias) in the astrometric radial velocity derived from the 
model fitting.  Such components of $\vec{T}$ are `observable' and in principle 
not harmful to the method.  Let us now examine more generally the extent to 
which $\vec{T}$ is observable by astrometry.

Suppose there exists a non-zero tensor $\vec{T}$ such that the velocity fields 
$\vec{v}_0 + \vec{T}\vec{s}$ and $\vec{v}_0 + \Delta\vec{v}$ produce identical 
observations for some vector $\Delta\vec{v}$.  Since the cluster velocity 
$\vec{v}_0$ is already a parameter of the model, the observational effects of 
the velocity field $\vec{T}$ could then entirely be absorbed by adjusting 
$\vec{v}_0$.  In this case $\vec{T}$ would be a non-observable component of the 
general velocity field.  Moreover, if there exists such a component in the real 
velocities, then the estimated cluster velocity will have a bias equal to 
$\Delta\vec{v}$.

We now need to calculate the effect of the arbitrary field $\vec{T}$  
on the observables.  Since the parallaxes are not affected, only the proper 
motion vector 
\begin{equation}
  \dot{\vec{r}} = (\vec{I} - \vec{r}\vec{r}^\prime)\vec{v}\pi/A
\end{equation}
needs to be considered.  In this equation $\vec{r}$ is the unit vector from the 
Sun towards the star, $\dot{\vec{r}}$ is the rate of change of that direction, 
and $\vec{I} - \vec{r}\vec{r}^\prime$ is the tensor representing projection 
perpendicular to $\vec{r}$ [$\vec{I}$ is the unit tensor; thus $(\vec{I} - 
\vec{r}\vec{r}^\prime)\vec{x}= \vec{x}-\vec{r}\vec{r}^\prime\vec{x}$ is the 
tangential component of the general vector $\vec{x}$].  With $b=|\vec{b}|=A/\pi$ 
we can write $\vec{s} = \vec{r}b - \vec{b}_0$.  $\vec{T}$ is non-observable if 
the space velocities $\vec{v}_0 + \vec{T}\vec{s}$ and $\vec{v}_0 + 
\Delta\vec{v}$ produce identical tangential velocities for every star, i.e.\ if
\begin{equation}
  (\vec{I} - \vec{r}\vec{r}^\prime)[\vec{v}_0+\vec{T}(\vec{r}b - \vec{b}_0)]
  = (\vec{I} - \vec{r}\vec{r}^\prime)(\vec{v}_0+\Delta\vec{v})
\end{equation}
for all directions $\vec{r}$ and distances $b$.  This is equivalently written
\begin{equation}
  (\vec{I} - \vec{r}\vec{r}^\prime)\vec{T}\vec{r}b - 
  (\vec{I} - \vec{r}\vec{r}^\prime)(\vec{T}\vec{b}_0+\Delta\vec{v}) = \vec{0} 
\, .
\end{equation}
In order that this should be satisfied for all $b$, it is necessary that  
$(\vec{I} - \vec{r}\vec{r}^\prime)\vec{T}\vec{r} = \vec{0}$ and $(\vec{I} - 
\vec{r}\vec{r}^\prime)(\vec{T}\vec{b}_0+\Delta\vec{v}) = \vec{0}$ are 
separately satisfied for all unit vectors $\vec{r}$.  The latter equation  
implies that
\begin{equation}
  \Delta\vec{v} = -\vec{T}\vec{b}_0 \, .
\label{eq:deltav}
\end{equation}
The former equation can be written $\vec{T}\vec{r} = 
\vec{r}\vec{r}^\prime\vec{T}\vec{r}$, which shows that $\vec{r}$ is an  
eigenvector of $\vec{T}$ (with eigenvalue $\vec{r}^\prime\vec{T}\vec{r}$).  
But the only tensor for which every unit vector is 
an eigenvector is the isotropic tensor, $\vec{T}=\vec{I}\kappa$ for the arbitrary 
scalar $\kappa$.  It follows that the only non-observable component of $\vec{T}$ is 
of the form $\vec{I}\kappa$, parametrised by the single scalar $\kappa$, and that 
consequently eight linearly independent components of $\vec{T}$ can, in 
principle, be determined from the astrometric observations.  The non-observable 
field $\vec{T}=\vec{I}\kappa$ describes a uniform isotropic expansion ($\kappa>0$) or 
contraction ($\kappa<0$) of the cluster with respect to its centroid.  These effects 
are observationally equivalent to an approach or recession of the cluster, 
i.e.\ to a different value of its radial velocity.

$\Delta\vec{v}$ is the bias for the centroid velocity. For any given star, the 
bias vector $\delta\vec{v}$ is the difference between the derived (apparent) 
space velocity vector $\vec{v}_0+\Delta\vec{v}$ and the true vector 
$\vec{v}_0+\vec{T}\vec{s}$.  Using $\vec{s}=\vec{b}-\vec{b}_0$ and 
Eq.~(\ref{eq:deltav}) we find
\begin{equation}
  \delta\vec{v} = -\vec{T}\vec{b} \, .
  \label{eq:deltav2}
\end{equation}
The resulting bias in the astrometric radial velocity is
\begin{equation}
  \delta(v_r) = -\vec{r}^\prime\vec{T}\vec{b} \, .
  \label{eq:deltavr}
\end{equation}
Isotropic expansion ($\vec{T}=\vec{I} \kappa$), in particular, gives the bias
\begin{equation}
  \delta_{\rm exp}(v_r) = -b\kappa \, .
  \label{eq:bias}
\end{equation}
For a uniformly expanding cluster $\kappa^{-1}$ equals the expansion age, i.e.\ the
time elapsed since all the stars were confined to a very small region of space.

\end{document}